\documentclass[]{epl}

\title{Subcritical crack growth in fibrous materials.}

\shorttitle{Subcritical crack growth in fibrous materials}

\author{S. Santucci\inst{1} \and P.-P. Cortet\inst{1} \and S. Deschanel\inst{1, 2} \and L. Vanel\inst{1}, \and S. Ciliberto\inst{1}.}
\shortauthor{S. Santucci \etal} \institute{ \inst{1}Laboratoire
de Physique, CNRS UMR 5672, Ecole Normale Sup\'erieure de Lyon \\
46 all\'ee d'Italie, 69364 Lyon Cedex 07, France\\ \inst{2}
GEMPPM, INSA de Lyon, 20 Av. Albert Einstein, 69621 Villeurbanne
Cedex, France}


\begin{document}

\maketitle

\begin{abstract}
We present experiments on the slow growth of a single crack in a
fax paper sheet submitted to a constant force $F$. We find that
statistically averaged crack growth curves can be described by
only two parameters : the mean rupture time $\tau$ and a
characteristic growth length $\zeta$. We propose a model based on
a thermally activated rupture process that takes into account the
microstructure of cellulose fibers. The model is able to reproduce
the shape of the growth curve, the dependence of $\zeta$ on $F$ as
well as the effect of temperature on the rupture time $\tau$. We
find that the length scale at which rupture occurs in this model
is consistently close to the diameter of cellulose microfibrils.
\end{abstract}

\section{Introduction}

A critical stress threshold is often defined to characterize
material resistance to rupture. When the applied stress is smaller
than the threshold, experiments show that rupture can still occur
after a delay in time which decreases with the applied stress and
with temperature \cite{Brenner,Zhurkov,Pauchard}. A widespread
approach has been to relate time-dependent fracture to the creep
properties of the material \cite{Schapery86,Kaminskii2}. In this
general framework, the creep law is an empirical material
dependent property, the most common dependence on the applied
stress proposed in the literature being either a power law or an
Arrhenius law. More recently, several authors have come back to a
simpler situation where the material is elastic and subcritical
rupture is thermally activated. The focus has been mainly on
prediction of the rupture time (or lifetime)
\cite{Sethna,Golubovic,Pomeau1}, but has also been extended to the
slow growth of a single crack in brittle materials \cite{Marder,
Santucci1}. In this Letter, we present experimental results on the
slow growth of a single crack in a sheet of paper which behaves as
a quasi-brittle material. We show that the experimental results
are compatible with a model of thermally activated crack growth in
brittle materials when we take into account the microstructure of
cellulose fibers in paper.

\section{Experimental set-up}

\begin{figure}[h!]
\centerline{\includegraphics[height=5cm]{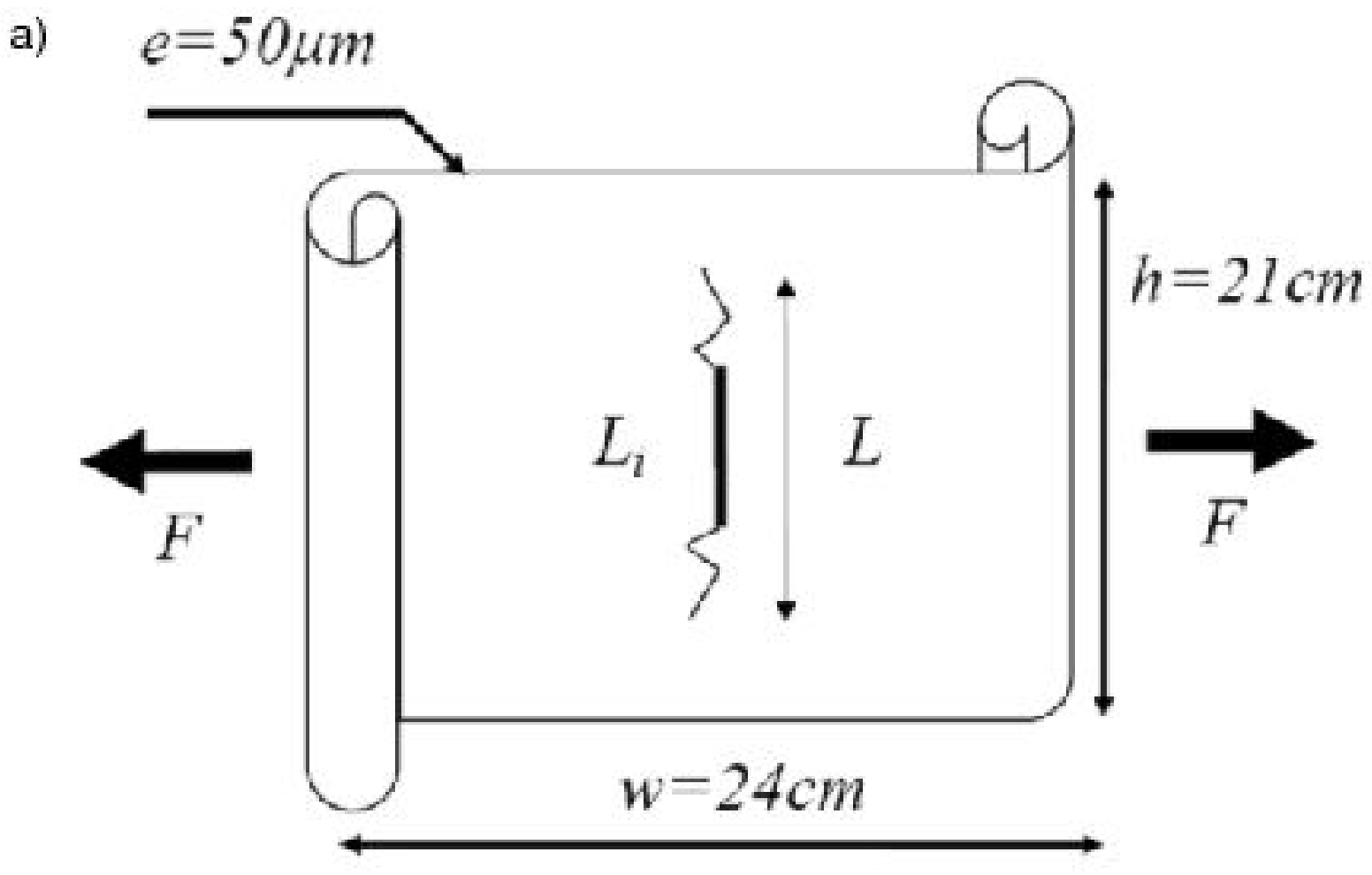}
\includegraphics[height=5cm]{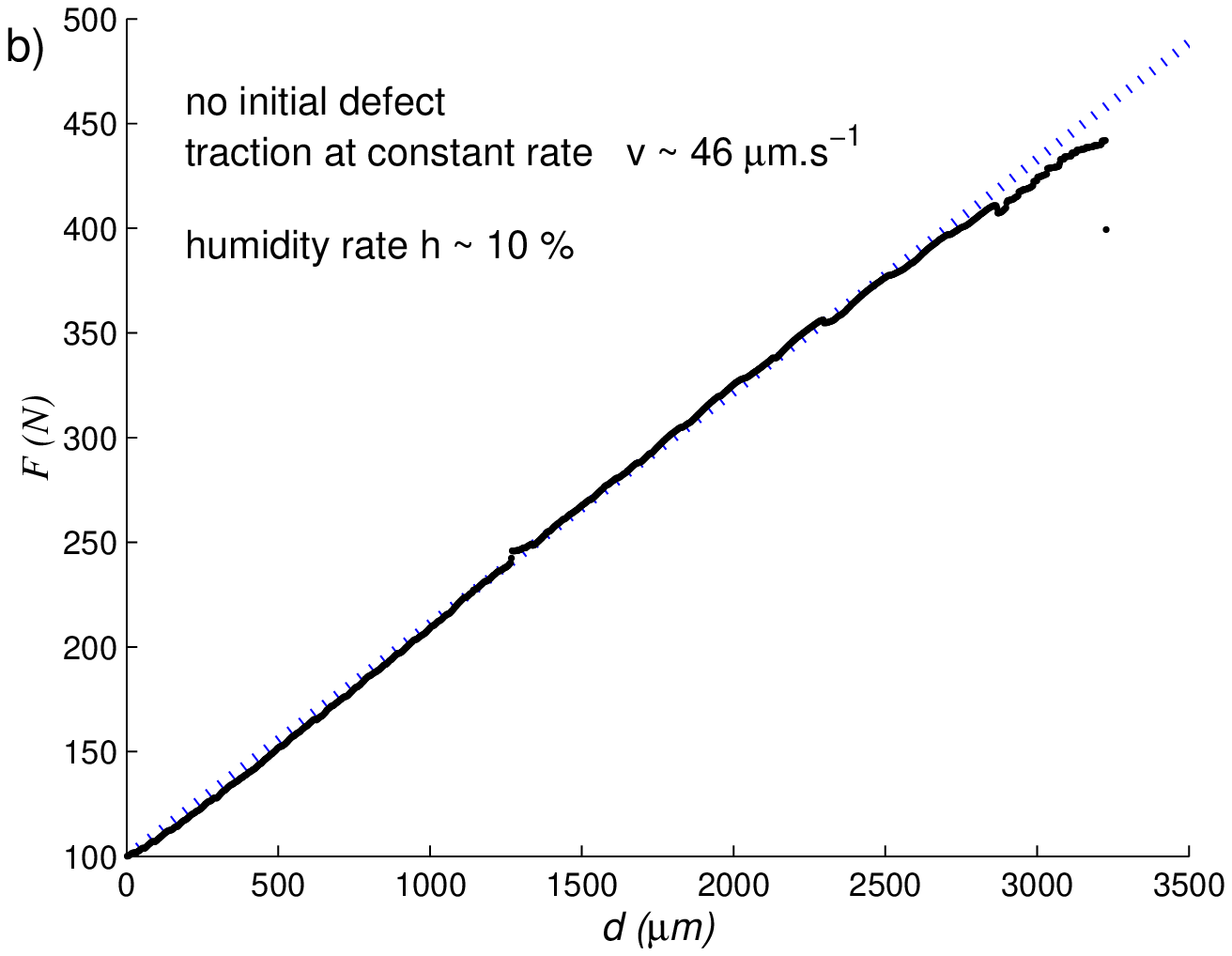}}
\caption{a) Sample geometry. b) Linear dependence between applied
force and elongation until rupture.} \label{f.1}
\end{figure}

Crack growth is obtained by loading in mode 1 at a constant force
$F$ a sheet of fax paper (Alrey) with an initial crack in the
center (fig.~\ref{f.1}a). The sample dimensions are : height $h =
21 \rm{cm}$, width $ w=24\rm{cm}$, and thickness $e = 50 \rm{\mu
m}$. The experimental set-up consists of a tensile machine driven
by a motor (Micro Controle $UE42$) controlled electronically to
move step by step (Micro Controle $ITL09$). The paper sheets are
mounted on the tensile machine with both ends attached with glue
tape and rolled twice over rigid bars clamped on jaws. The motor
controls the displacement of one jaw ($400$ steps per micrometer)
while the other jaw is rigidly fixed to a force gage
(Hydrotonics-TC). The tensile machine is placed in a box with a
humidity level stabilized at $5\%$. In order to work on samples
with the same initial crack shape and length $L_i$, we use
calibrated razor blades mounted on a micrometric screw and we
initiate a macroscopic crack precisely at the center of the sheet.
The samples are loaded by increasing the distance between the jaws
such that the resulting force $F$ is perpendicular to the initial
crack direction.

A feedback loop allows us to adjust the displacement in order to
keep the applied force $F$ constant with a precision better than
$0.5 \rm{N}$ and a response time less than $10 \rm{ms}$. From the
area $A$ of a cross-section of the sheet, $A$ being
approximatively constant, we calculate the applied stress
$\sigma_e=F/A$.

\section{Physical properties of paper}

Sheets of fax paper break in a brittle manner. This is evidenced
by the elastic stress-strain dependence which is quasi-linear
until rupture (fig.~\ref{f.1}b). Another sign that rupture is
essentially brittle is given by the very good match between the
two opposite lips of the fracture surfaces observed on post-mortem
samples.

A sheet of paper is a complex network of cellulose fibers.
Scanning electron microscopy on our samples shows fiber diameters
between $4 $ and $50 \rm{\mu m}$ with an average of $18 \rm{\mu
m}$. Cellulose fibers are themselves a bundle of many
microfibrils. Cellulose microfibrils have a cristalline structure
(therefore, they are very brittle) and are consistently found to
have a diameter $d=2.5 nm$ \cite{Jakob}.

The mechanical properties of paper depend crucially on the
humidity rate. To get reproducible results, the fax paper samples
are kept at least one day at a low humidity level ($< 10 \%$) and
during the experiment ($\simeq 5\%$). At constant humidity level
($h \simeq 5\%$) and room temperature, the Young modulus of the
fax paper sheets is typically $Y = 3.3\,10^9 \rm{N.m^{-2}}$.

\section{Direct observation and image analysis}

We light the samples from the back. A high resolution and high
speed digital camera (Photron Ultima 1024) collects the
transmitted light and allows us to follow the crack growth. We
observe that the global deformation of the paper sheet during a
creep experiment is correlated in a rather reproducible way to the
crack growth whatever the rupture time. We use this property to
trigger the camera at fixed increment of elongation (one micron)
rather than at fixed increment in time. This avoids saturation of
the onboard memory card when the crack growth is slow and makes
the acquisition rate faster when the crack grows faster and starts
to have an effect on global deformation. We acquire $2$ frames at
$250 \rm{fps}$ at each trigger and obtain around one thousand
images per experiment.

\begin{figure}[h!]
\onefigure[width=12cm]{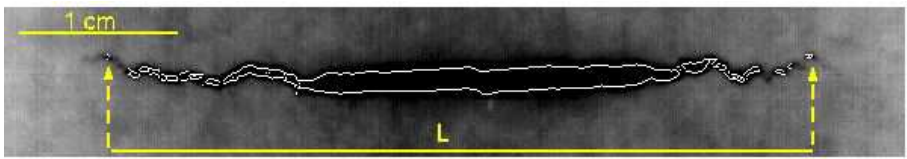} \caption{ Extraction of the
projected crack length $L$ from the crack contour detected.}
\label{extract_L}
\end{figure}

Image analysis is performed to extract the length of the crack
projected on the main direction of propagation, i.e. perpendicular
to the direction of the applied load (fig.~\ref{extract_L}).
Although the crack actually follows a sinuous trajectory, its
projected length $L$ gives the main contribution to the stress
intensity factor $K$ which we compute as: $K=g(\pi L/2 h) \sigma_e
\sqrt{\pi L/2}$, where $g(x)=\sqrt{\tan x/x}$ takes into account
the finite height $h$ of the paper sheet \cite{Lawn}.

\section{Single crack growth}
For a given initial crack length $L_i$, subcritical crack growth
is obtained by applying a constant force $F$ so that $K(L_i)$ is
smaller than the critical rupture threshold of the material $K_c$
above which fast crack propagation would occur. During an
experiment, the crack length increases, and so does the stress
intensity factor $K(L)$. This will cause the crack to accelerate
until it reaches the critical length $L_c$ for which $K(L_c)=K_c$.

\begin{figure}
\centerline{\includegraphics[width=7cm]{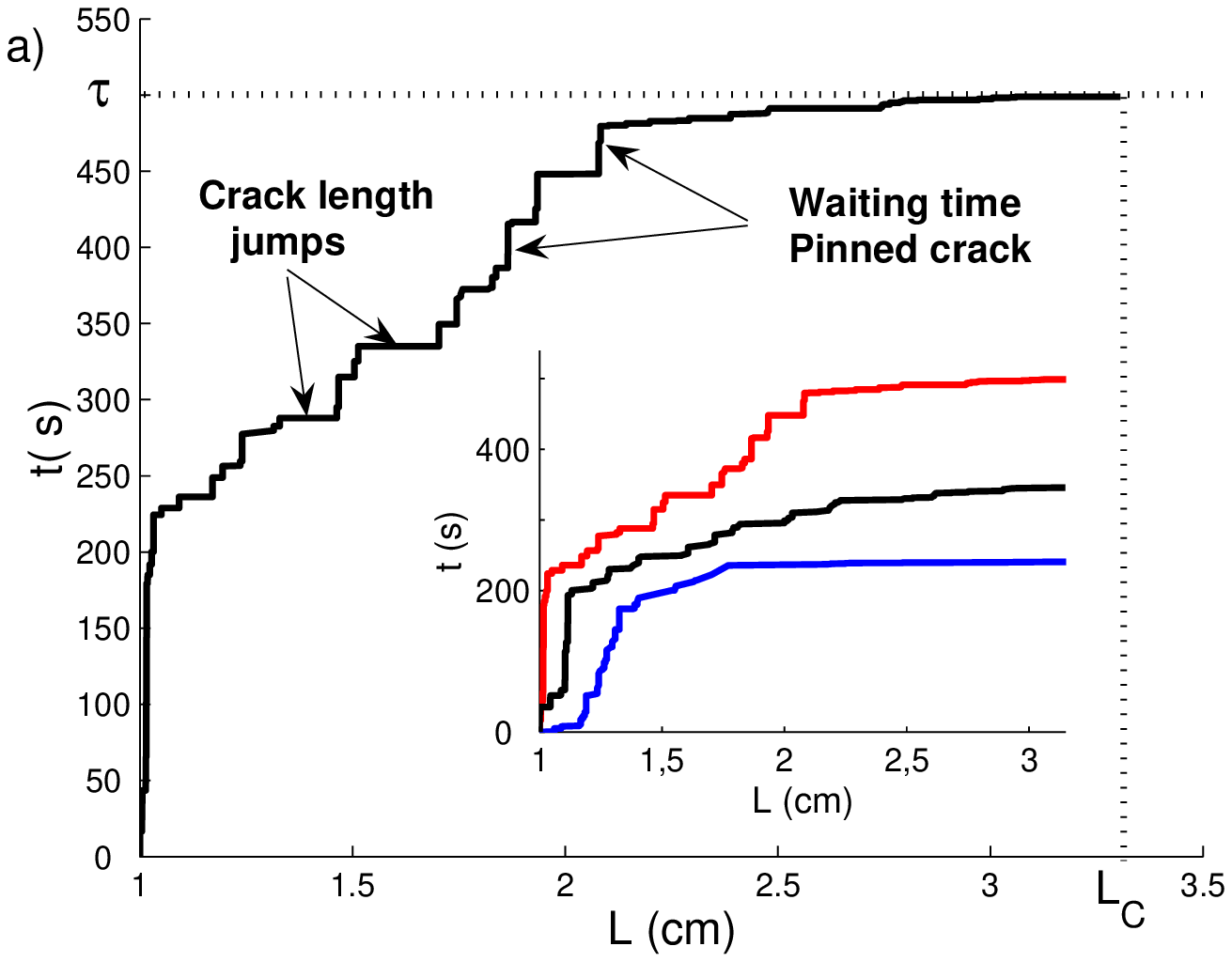}
\includegraphics[width=7cm]{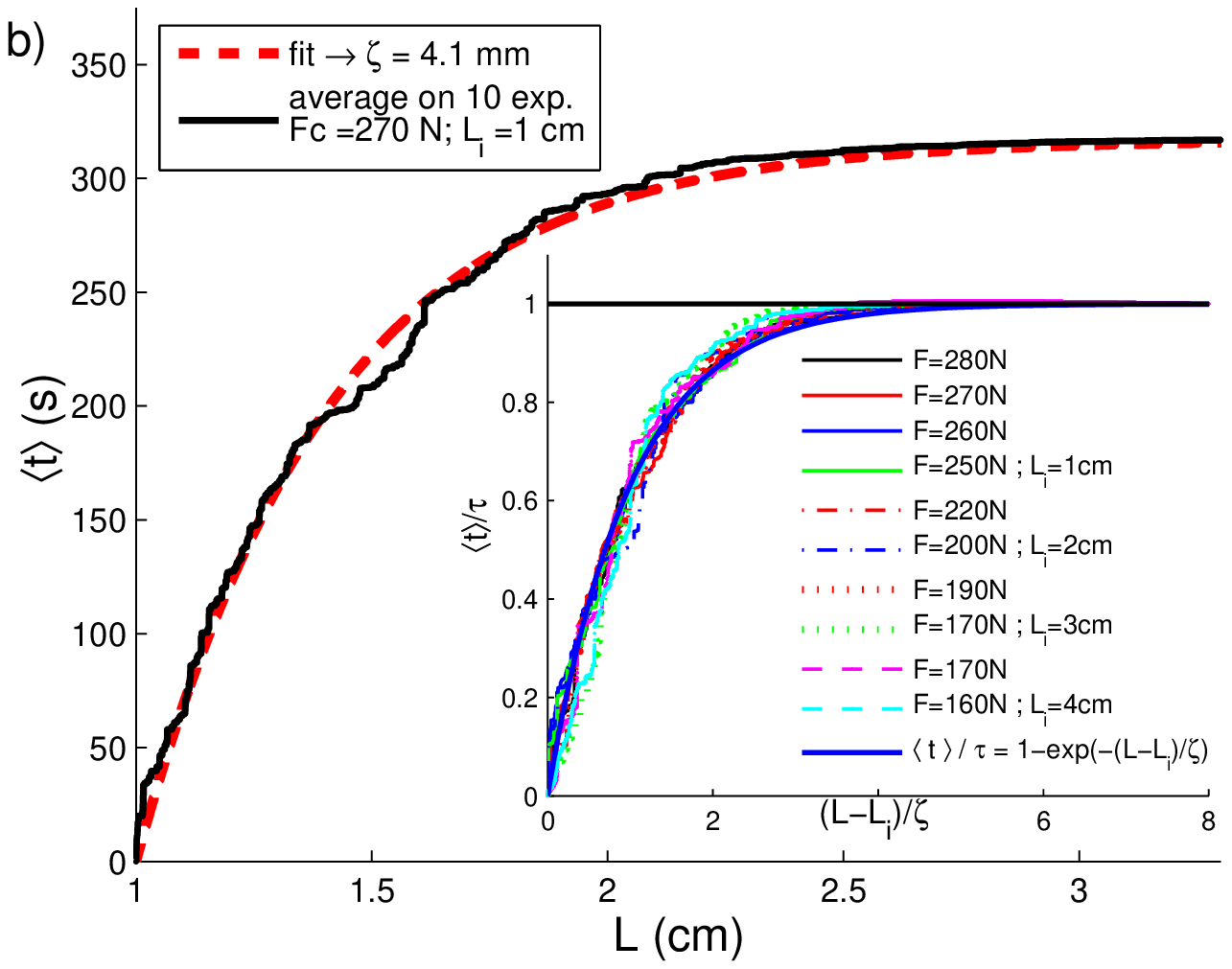}}
\caption{ a) Typical stepwise growth curve for a creep experiment
with an initial crack length $L_i=1\rm{cm}$ submitted to a
constant load $F=270\rm{N}$. The lifetime of the sample is $\tau
=500\rm{s}$ and the critical length $L_c=3.3\rm{cm}$. In insert, a
strong dispersion is observed in crack growth profile and in
lifetime for $3$ creep experiments realized in the same
conditions. b) Statistical average of growth curves for $10$ creep
experiments realized in the same conditions ($L_i=1\rm{cm}$,
$F=270\rm{N}$). The dashed line corresponds to a fit using
equation eq.~(\ref{eq.growth}) with a single free parameter $\zeta
= 0.41 \rm{cm}$. Insert: rescaled average time $\langle t
\rangle/\tau$ as function of rescaled crack length $(L-L_i)/\zeta$
for various initial crack lengths and applied stresses. The solid
line corresponds to eq.~(\ref{eq.growth}).} \label{growth}
\end{figure}

 On fig.~\ref{growth}a we show a typical growth curve
obtained during a creep experiment with an applied force
$F=270\,\rm{N}$ and an initial crack length $L_i=1\rm{cm}$. Since
time to rupture $\tau$ is a statistical quantity, we prefer to
plot time evolution as a function of the crack length. We observe
that the crack growth is actually intermittent. Essentially, there
are periods of rest during which the crack tip is pinned and does
not move, and also moments when the crack suddenly opens and
advances of a certain step size $s$. The crack advances by jumps
until it reaches a critical length $L_c$ where the paper sheet
breaks suddenly. Measurements of $L_c$ are used to estimate the
critical stress intensity factor $K_c=6 \pm
0.5\,\rm{MPa.m^{1/2}}$. Beyond $L_c$, the crack runs across the
whole sample (about $18 \rm{cm}$ in this case) in less than one
second, with a crack speed $v > 5 \rm{m.s^{-1}}$. For the same
experimental conditions (same stress, same initial crack length,
same temperature and same humidity rate), we observe a strong
dispersion in growth curves and in lifetime  while the critical
length seems to be rather well defined (see insert in
fig.~\ref{growth}). In order to characterize both the average
crack growth and the stepwise growth dynamics, a statistical
analysis is required. In this Letter, we are going to focus our
study on the average dynamics. The intermittent dynamics and in
particular the step size statistics has already been described
elsewhere \cite{Santucci2}.

\section{Statistically averaged crack growth}
We have performed an extensive study of crack growth, varying the
initial crack length from $L_i=1 \rm{cm}$ to $L_i=4\rm{cm}$ and
the applied force between $F=140 \rm{N}$ and $F=280\rm{N}$
(corresponding to an initial stress intensity factor between
$K_i=2.7\rm{MPa.m^{1/2}}$ and $K_i=4.2\rm{MPa.m^{1/2}}$) and
repeating $5$ to $20$ experiments in the same conditions (stress,
initial crack length, temperature and humidity rate). The
resulting measured lifetime varied from a few seconds to a few
days depending on the value of the applied stress or the
temperature. In order to characterize the average growth dynamics,
we examine for given experimental conditions the average time
$\langle t \rangle (L)$ the crack takes to reach a length $L$.

Even though the lifetime distribution is large and the growth
dynamics intermittent, the average growth offers a regular
behavior. A typical mean growth, obtained by averaging ten
experiments in the same conditions, is plotted in
fig.~\ref{growth}b. This growth is qualitatively very close to an
exponential evolution:
\begin{equation}\label{eq.growth}
\langle t \rangle = \tau \left[1 -
\exp\left(-\frac{L-L_i}{\zeta}\right)\right]
\end{equation}

Indeed, we obtain a very good fit of the data in
fig.~\ref{growth}b with eq.~(\ref{eq.growth}) setting the mean
lifetime to the experimentally measured value and using $\zeta$ as
a unique free parameter. Using the same procedure, we extract the
characteristic growth length $\zeta$ for various experimental
conditions. In the insert of fig.~\ref{growth}b, rescaling the
crack length by $\zeta$ and the time by $\tau$ for many different
experimental conditions, we show that the data collapse on the
functional form given by eq.~(\ref{eq.growth}). Moreover, we have
checked that the deviation from the predicted average behaviour is
smaller when increasing the number of experiments.

\begin{figure}[h!]
\centerline{\includegraphics[width=7cm]{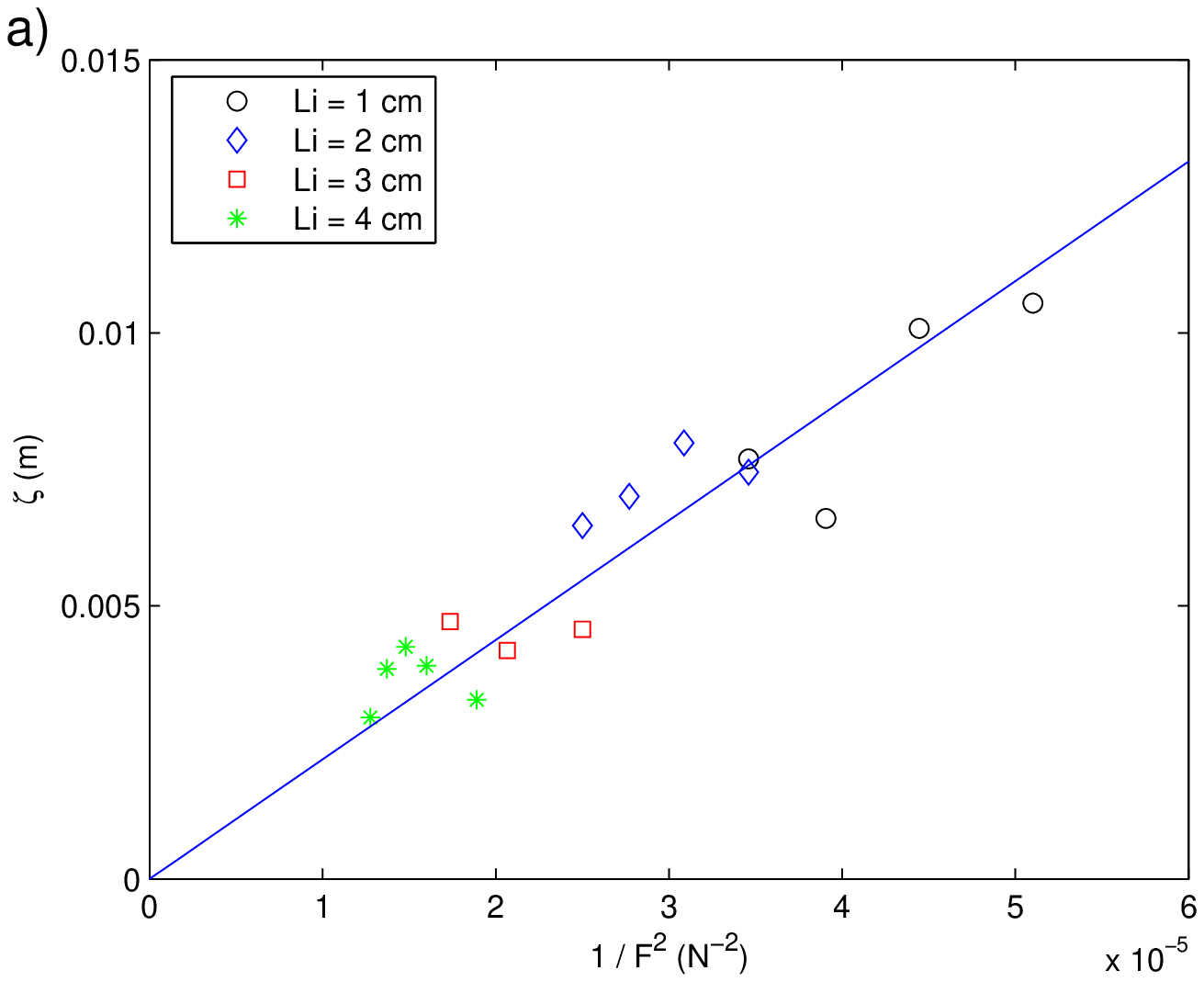}
\includegraphics[width=7cm]{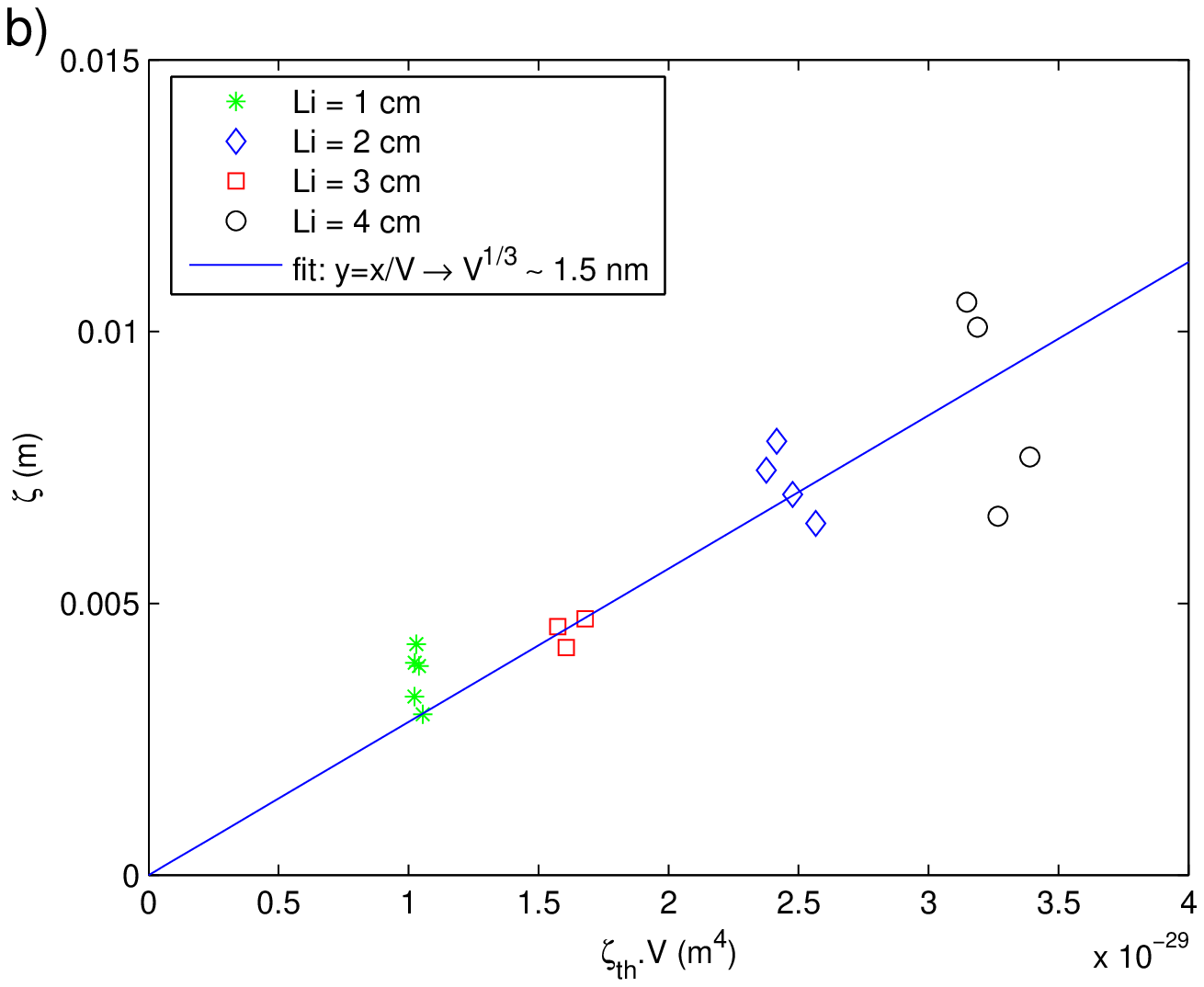}}
\caption{ a) Experimental value of $\zeta$ extracted from the
average growth profile as a function of $1/F^2$. b) $\zeta$ as a
function of the prediction of the thermally activated rupture
model. The line represents the best linear fit $y=x/V$. Its slope
permits us to obtain a characteristic length scale for rupture:
$V^{1/3}\sim 1.5\,\rm{nm}$.} \label{zeta}
\end{figure}

\begin{figure}
\centerline{\includegraphics[width=7cm]{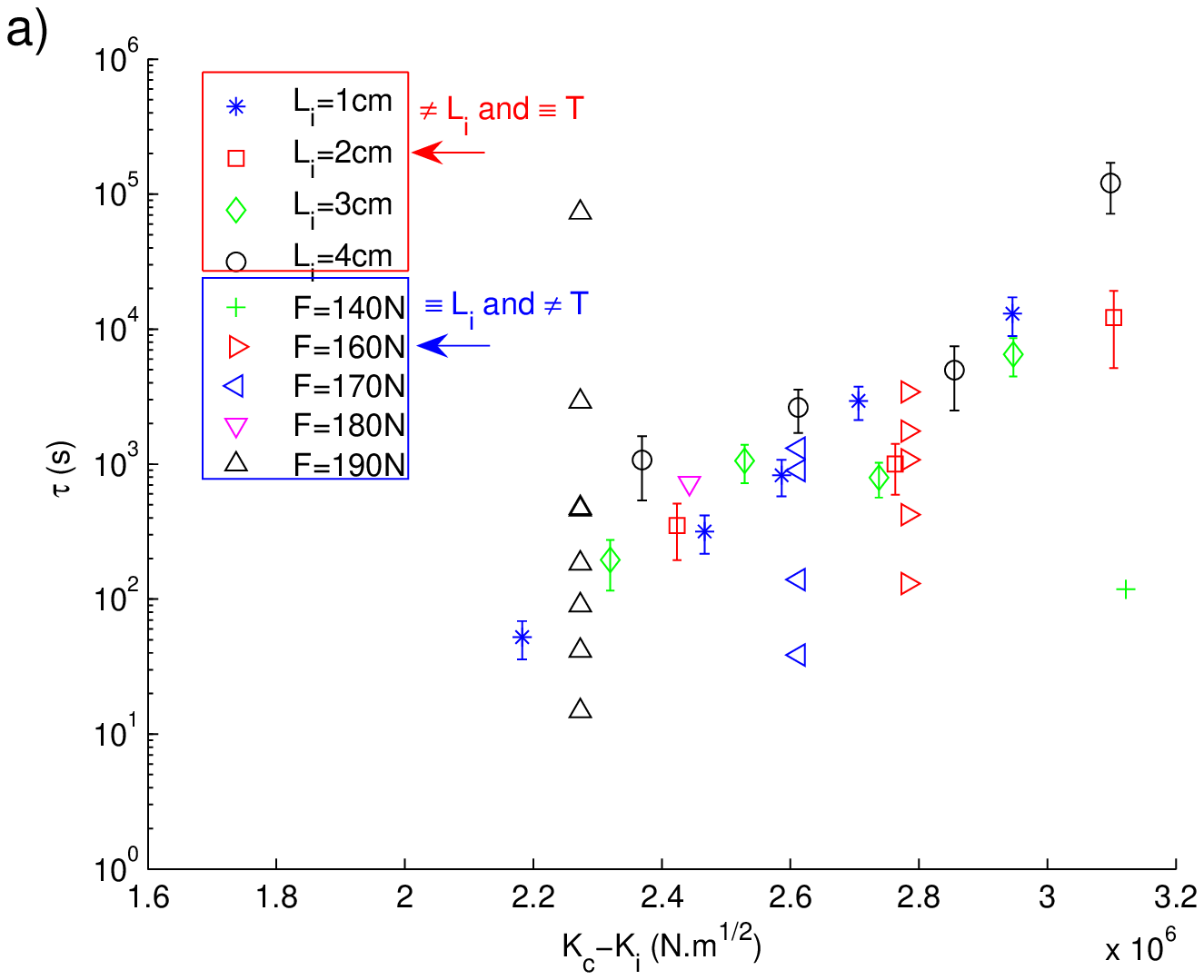}\includegraphics[width=7cm]{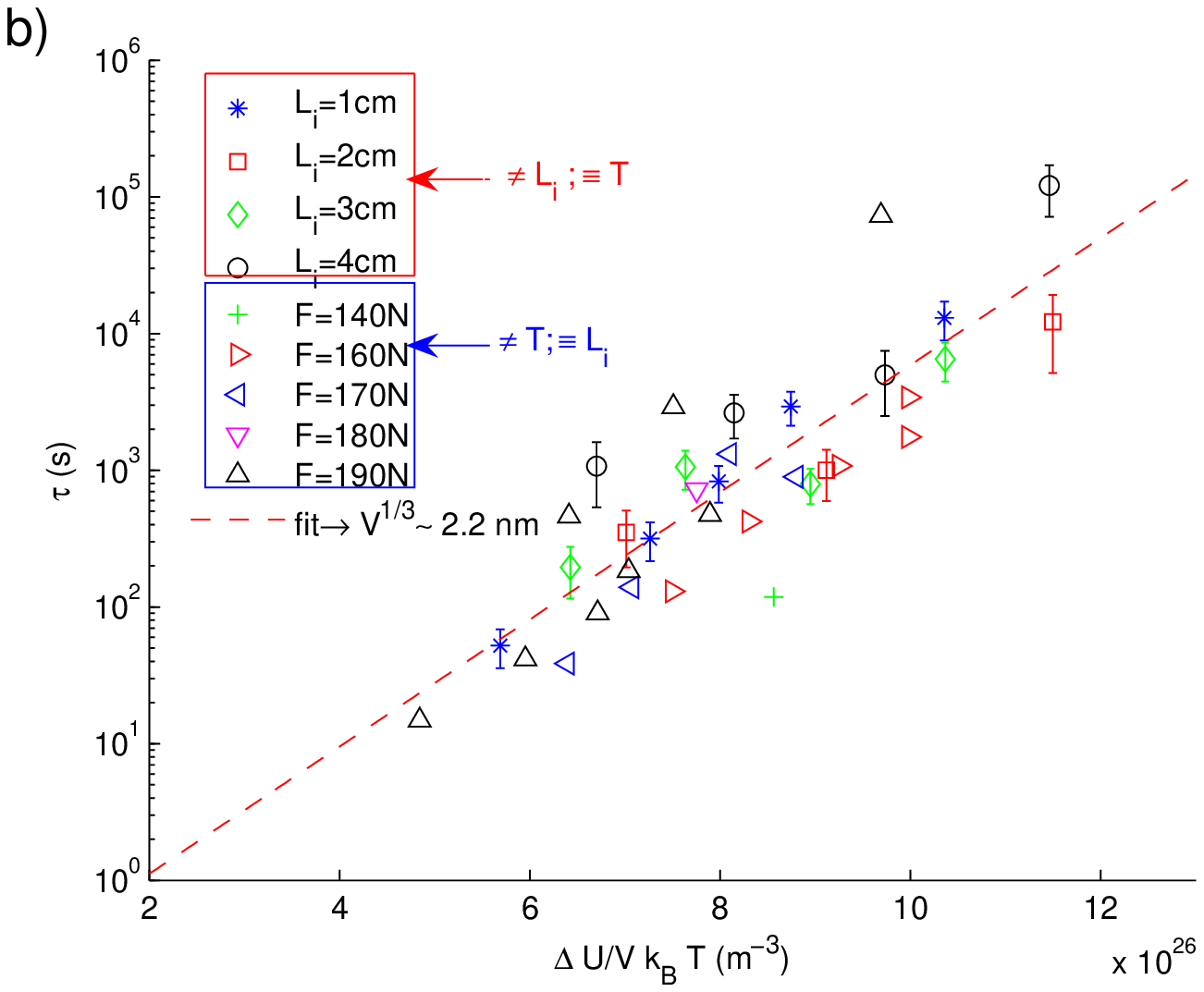}}
\caption{a) Logarithm of the mean rupture time $\tau$ as a
function of $K_c-K_i$. Data points without error bars correspond
to non-averaged measurements obtained by varying temperature with
$L_i=2\rm{cm}$ and various fixed values of $F$. b) Logarithm of
lifetimes as a function of the dimensional factor $\Delta U/V k_B
T$ predicted by eq.~(\ref{eqtau}) for different values of $L_i$,
$F$ and $T$. The best fit $\log\,\tau\,\propto\,\Delta U / k_B T$
(dashed line) slope gives an estimation of the characteristic
length scale $V^{1/3} \sim 2.2 \rm{nm}$.} \label{fig.Lifetime}
\end{figure}

We find that the experimental value of $\zeta$ is approximatively
proportional to $1/F^2$ (fig.~\ref{zeta}a). We also observe that
the rupture time for all the experiments performed at room
temperature is essentially a function of the initial stress
intensity factor $K_i$ (symbols with an error bar on
fig.~\ref{fig.Lifetime}a. However, for a fixed value of applied
force $F$ and initial length $L_i$, varying temperature between
$20^{\circ}\rm{C}$ and $120^{\circ}\rm{C}$ leads to variations of
the rupture time up to four order of magnitude (symbols without
error bar on fig.~\ref{fig.Lifetime}a).

\section{Model of thermally activated crack growth}

In this part, we are going to describe a model that is able to
reproduce several of the experimental features that we have just
described.

First, we recall that the experimental crack growth curves have
been obtained by performing a statistical average on many
experiments in the same condition. The model will also be a
statistical one. We are going to consider that in average the
crack growth proceeds by breaking individual cellulose fibers at
the crack tip. Still in average, the fibers who resist to the
applied load will be considered parallel to each other and
perpendicular to the crack direction. If $\lambda$ is the typical
diameter of a fiber and $\tau_L$ is the mean rupture time for the
cellulose fiber at the crack tip, the mean crack velocity is :
$v=\lambda / \tau_L $. In that way, the increase in crack velocity
as the crack length increases will correspond to a decrease of the
rupture time $\tau_L$.

Whatever is the rupture mechanism, we expect that the rupture time
$\tau_L$ is going to depend on the applied load $f$ on the fiber
at the crack tip. Equivalently, we can refer to the applied stress
$\sigma_L = f / S$ where $S \sim \lambda^2$ is the cross-sectional
area of the fiber. We estimate $\sigma_L$ assuming that the fiber
acts as a cohesive zone of length $\lambda$ and that the rest of
the material behaves as an elastic continuum for a crack with a
length $L+\lambda$. Following \cite{Barenblatt}, we consider that
the crack is at equilibrium if the elastic field presents no
divergence of stress. This condition imposes that the stress level
in the fiber is controlled by the size of the fiber and can be
estimated as: $\sigma_L \propto \sigma_e
\sqrt{L+\lambda}/\sqrt{\lambda}\sim \sigma_e \sqrt{L/\lambda}$. In
our calculations, we will take $\sigma_L=K/\sqrt{2\pi\lambda}$
where $K$ is the stress intensity factor.

A common approach to describe the creep rupture of materials is to
introduce a time-dependent creep compliance or a rate of rupture
which is a power law of the applied stress. Instead of assuming a
phenomenological law for the creep behavior, we are going to use a
statistical approach that has been developed in the case of
brittle materials to describe the rupture of fiber bundles. This
approach is appropriate in the case of a cellulose fiber because
it is actually a bundle of many identical brittle microfibrils.
The idea of the rupture mechanism is that at equilibrium there are
always statistical fluctuations of stress with a variance
depending on the actual temperature of the material. Taking into
account the probability that such stress fluctuations exceed the
rupture threshold $f_c$ of individual fibers in the bundle, it is
possible to make some prediction on the rupture dynamics of a
fiber bundle. In an homogenous fiber bundle where all the fibers
have the same rupture thresholds, it was shown that the rupture
time is \cite{Ciliberto, Roux}:
\[\tau \sim \exp\left[\frac{(f_c-f_0)^2}{2 \alpha k_B T}\right]\]
where $k_B$ is Boltzmann constant, $T$ is temperature, $\alpha$ is
the elastic stiffness of the fibers, $f_0$ is the applied force
per fiber when the bundle is still intact, and $f_c$ is the
rupture threshold of the fibers. We can rewrite this expression in
term of the Young modulus $Y$, the applied stress on the fiber
$\sigma_L$ and the rupture threshold $\sigma_c$. If $s=\pi d^2/4$
is the section area of a microfibril, we have $f_0=\sigma_L s$ and
$f_c=\sigma_c s$. Rupture in a microfibril is expected to occur in
a small piece of the fiber with a length $\ell$ much smaller than
the total fiber length. The relation between the stiffness
$\alpha$ of this piece and the Young modulus is then: $\alpha=Y s
/\ell$. Introducing the volume $V=s \ell$, we see that the rupture
time can be written as :
\[\tau_L \sim \exp\left[\frac{(\sigma_c-\sigma_L)^2V}{2Y k_B T}\right]\]
It must be understood that in this formula $V$ represents the
volume in which rupture of the microfibril actually occurs. Since
the microfibril is broken whenever we cut a section through, it is
physically reasonable to consider that the length $\ell$ of the
piece involved in rupture will be comparable to the microfibril
diameter $d$. Thus, we will expect to have $ V \sim d^3$.

The expression obtained for $\tau_L$ gives an exponential
dependence of the crack growth velocity. If we integrate this
velocity following the procedure described in \cite{Santucci1}, we
recover eq.~(\ref{eq.growth}) where the rupture time is :
\begin{equation}\label{eqtau}
  \tau  \sim \exp \left( \frac{\Delta U}{k_B T}
  \right),\qquad \mathrm{with} \qquad \Delta U =\frac{(\sigma_c-\sigma_i )^2 V}{2 Y}
\end{equation}
and the characteristic growth length $\zeta$:
\begin{equation}\label{eqzeta}
  \zeta = \frac{2 Yk_BT}{V}
  \frac{L_i}{\sigma_i(\sigma_c-\sigma_i)},
\end{equation}
where $\sigma_i$ is the value of $\sigma_L$ when $L=L_i$.

Writing explicitly the dependence of stresses on $F$, we obtain
from eq.~(\ref{eqzeta}) that $\zeta \propto 1/F^2$ in agreement
with the experimental observations (fig.~\ref{zeta}a). We have
also $\zeta \propto \sqrt{L_i/(L_c-L_i)}$ but this dependence is
difficult to verify experimentally because the ratio $L_c/L_i$
does not change much. In fig.~\ref{zeta}b, we see that there is a
reasonable agreement between the experimental value of $\zeta$ and
the theoretical one $\zeta V = 2 Yk_BT L_i
/[\sigma_i(\sigma_c-\sigma_i)]$. Using $V$ as a free parameter, we
find a characteristic scale $V^{1/3}=1.5nm$ close to the
microfibril diameter $d$.

The form of the energy barrier $\Delta U$ in the Arrhenius law
(eq.~(\ref{eqtau})) is analogous to the one obtained for example
in \cite{Marder}. Note also that it is not simply a function of
the external load $\sigma_e$ as in
\cite{Sethna,Golubovic,Pomeau1}. In fig.~\ref{fig.Lifetime}b, we
plot the rupture time as a function of $\Delta U / V k_B T =
(\sigma_c-\sigma_i)^2/2 Y k_B T$. We observe that the data points
obtained by varying temperature are now collapsing with the one
obtained at fixed temperature (the dispersion in this figure
remains large because data as a function of temperature and
constant $F$ and $L_i$ have been averaged less than those at
constant temperature for technical reasons). From a fit of the
data, we obtain independently a new estimate of $V$ which gives a
characteristic scale $V^{1/3}=2.2 nm$. Once again this estimate is
close to the microfibril diameter $d$.

\section{Conclusion}
We have studied the subcritical growth of a single crack in a
sheet of paper during creep experiments. We have shown that
statistically the crack growth curves can be well described with
only two parameters: the rupture time $\tau$ and a characteristic
growth length $\zeta$. We have proposed a model which is able to
reproduce the shape of the growth curves as well as the dependence
of $\zeta$ with the applied load and the influence of temperature
on the rupture time. The model takes advantage of the fact
cellulose fibers are actually a bundle of many small brittle
fibers with a diameter of nanometer scale. A previous prediction
of the rupture time for thermally activated rupture of brittle
fibers has been used to derive the crack velocity. We find that
the experimental data are compatible with the idea that rupture
occurs at a nanometric scale in the cellulose fibers. While the
model we used work rather well, it remains to understand if the
experimental results could be as well described by time-dependent
fracture models based on the material rheological properties. Such
a study would require for example a detailed investigation of the
physical properties of individual fibers.

\acknowledgements This work was
partially supported by the Rh\^{o}ne-Alpes Emergence 2003 program.


\begin{thebibliography}{0}


\bibitem{Brenner}
  \Name{Brenner S. S.}
  \REVIEW{J. Appl. Phys.}{33}{1962}{33}.

\bibitem{Zhurkov}
  \Name{Zhurkov S. N.}
  \REVIEW{Int. J. Fract. Mech.}{1}{1965}{311}.

\bibitem{Pauchard} \Name{Pauchard L., \and Meunier J.}
  \REVIEW{Phys. Rev. Lett.}{70}{1993}{3565}.

\bibitem{Schapery86} \Name{Schapery R. A.}\Book{in Encyclopedia of Material Science and
Engineering}\Publ{Pergamon, Oxford} \Year{1986} \Page{5043}.

\bibitem{Kaminskii2}
  \Name{Kaminskii A. A.}
  \REVIEW{Int. Appl. Mech.}{40}{2004}{829}.

\bibitem{Sethna}
  \Name{Buchel A., \and Sethna J. P.}
  \REVIEW{Phys. Rev. Lett.}{77}{1996}{1520};
  \REVIEW{Phys. Rev. E}{55}{1997}{7669}.

\bibitem{Golubovic}
  \Name{Golubovic L., and Feng S.}
  \REVIEW{Phys. Rev. A}{430}{1991}{5233}.

\bibitem{Pomeau1}
  \Name{Pomeau Y.}
  \REVIEW{C.R. Acad. Sci. Paris II}{314}{1992}{553};\REVIEW{C.R. M\'ecanique}{330}{2002}{1}.
\bibitem{Marder}
 \Name{Marder M.}
 \REVIEW{Phys. Rev. E}{54}{1996}{3442}.

\bibitem{Santucci1} \Name{Santucci S., Vanel L., Guarino A., Scorretti R. \and Ciliberto S.}
\REVIEW{Europhys. Lett.}{62 (3)}{2003}{320}.
%
\bibitem{Jakob}\Name{Jakob H. F., Tschegg S. E. and Fratzl P.}\REVIEW{Struct. Biol.}{133}{1994}{13}.

\bibitem{Lawn} \Name{Lawn B. R. \and Wilshaw T. R.}
\Book{Fracture of Brittle Solids} \Publ{Cambridge University
Press, Cambridge} \Year{1975}.
\bibitem{Santucci2}
\Name{Santucci S., Vanel L., \and Ciliberto S.} \REVIEW{Phys. Rev.
Lett.}{93}{2004}{095505}.
\bibitem{Barenblatt} \Name{Barenblatt G. I.}\Book{in Advances in Applied
Mechanics}\Vol{VII}\Publ{Academic Press}\Year{1962}\Page{55}.

\bibitem{Roux}  \Name{Roux S.}\REVIEW{Phys. Rev. E}{62}{2000}{6164}.

\bibitem{Ciliberto}\Name{Ciliberto S., Guarino A., \and Scorretti R.}\REVIEW{Physica D}{158}{2001}
{83}.





\end{thebibliography}
\end{document}